\documentclass[sigconf, nonacm]{acmart}

\usepackage{listings}
\usepackage{balance}

\AtBeginDocument{%
  }

\begin{document}

\title{``CS 1.5'': An Experience Report on Integrating CS1 and Discrete Structures for the AI Era}

\author{Ildar Akhmetov}
\email{i.akhmetov@northeastern.edu}
\orcid{0000-0002-6660-8890}
\affiliation{%
  \institution{Northeastern University}
  \city{Vancouver}
  \state{BC}
  \country{Canada}
}

\author{Juancho Buchanan}
\email{j.buchanan@northeastern.edu}
\affiliation{%
  \institution{Northeastern University}
  \city{Vancouver}
  \state{BC}
  \country{Canada}
}

\renewcommand{\shortauthors}{Akhmetov and Buchanan}

\date{March 2026}

\begin{abstract}
The rapid proliferation of generative AI has fundamentally altered the landscape of introductory computer science education. Traditional methods that prioritize syntax memorization and writing code from scratch are challenged by tools that can generate such code instantly. In response, we designed and implemented an experimental course integration at Northeastern University Vancouver, merging ``Intensive Foundations of Computer Science'' (CS1) and ``Discrete Structures'' into a single, cohesive studio experience. Dubbed ``CS 1.5''---a playful nod to its position between CS1 and CS2---this course operates on two core principles: embracing AI as a collaborator rather than an adversary, and prioritizing deep theoretical foundations alongside practical implementation. This report details our pedagogical interventions, including the restructuring of the timetable to support a 4-hour studio format, the introduction of ``sharing circles'' to foster human connection, and the strategic shift to ``code comprehension'' over code generation. We discuss specific integrated projects—spanning set theory, recursion, and probability—that bridge the gap between mathematical proofs and software implementation. Finally, we reflect on the changing role of the instructor—from a repository of knowledge to a human mentor—and offer practical recommendations for scaling this high-touch, integrated model.
\end{abstract}


\maketitle

\section{Introduction}

The advent of capable Large Language Models (LLMs) has forced a reckoning in computer science education. The traditional CS1 curriculum—often focused on syntax memorization, basic control structures, and writing small programs from scratch—faces existential challenges in a world where AI can generate such code instantly \cite{prather2023we, becker2023programming}. Both authors, teaching in the Align Bridge program at Northeastern University Vancouver (a program designed to transition non-CS majors into a Master of Science in Computer Science), agreed that we could not teach in 2025 the way we did in 2015. We observed that assignments designed to test logical thinking were increasingly becoming tests of prompt engineering, and students were often bypassing the "struggle" necessary for building mental models. We needed a bold experiment.

The inspiration for our approach came during a long run, where one of the authors realized the inherent natural unity between the two courses we were teaching: \textit{CS 5001: Intensive Foundations of Computer Science} and \textit{CS 5002: Discrete Structures}. Concepts like recursion and induction, or Boolean algebra and conditional statements, are not separate disciplines but mirrors of each other. Yet, they are traditionally taught in silos, often in different semesters or by different instructors who rarely coordinate.

In the Fall of 2025, we launched a pilot with a cohort of 11 students to merge these courses into a single experimental offering. Among ourselves we called it ``CS 5001.5''---a playful portmanteau of the two course numbers---but as we began preparing materials for open-source release, the shorter ``CS 1.5'' stuck as its public identity, a nod to the course's position between CS1 and CS2. By manipulating the timetable to schedule the courses back-to-back, we created a 4-hour studio block (plus a 2-hour weekly recitation), giving us the room to experiment with a ``flipped classroom'' and ``studio-based'' learning model.

Our goal was twofold: to find a way to teach Computer Science in the era of AI, and to ensure students built an intuition for problem-solving that transcends tools. We operated on two core principles:
\begin{enumerate}
    \item \textbf{AI Embraced:} Rather than banning LLMs, we integrated them as core components of the class, shifting the focus from code generation to code comprehension and system design. This was operationalized by assigning large, complex codebases that required AI assistance to navigate.
    \item \textbf{Foundations First:} We utilized paper-based exercises and physical props to ensure students grasped the underlying logic before delegating implementation to the machine. This was operationalized through activities like ``dancing bubble sort'' and manual proofs of recursive costs.
\end{enumerate}

This experience report details the design, execution, and outcomes of this experimental course.

\section{Context: The Align Program}

The Align Bridge program serves a unique demographic: students with bachelor's degrees in non-computing fields (ranging from Philosophy and English to Biology and Music) who are transitioning into computer science. Align students hold undergraduate degrees across 23 different fields—roughly 55\% from STEM disciplines (primarily Physical Sciences and Engineering) and 45\% from non-STEM fields (primarily Liberal Arts, Business, and Social Sciences) \cite{Schmidt2025AnMS}. This diversity became a central asset in our ``sharing circles'' and narrative-based pedagogy, as students could draw on their varied life experiences to understand abstract CS concepts.

For this specific experimental cohort, we worked with a small group of 11 students. While students in the standard curriculum typically take both CS1 and Discrete Structures simultaneously, the courses are traditionally taught as distinct silos with little to no coordination. This context—a highly motivated, mature, and diverse student body taking both courses in parallel—shaped our decision to attempt a full integration. We knew we could rely on their prior academic maturity to navigate the experimental nature of the merged curriculum.

\section{Related Work}

Recent work in computing education has highlighted how generative AI tools are reshaping what it means to ``learn programming.'' Studies of AI code-generation tools with novices report both productivity gains and new failure modes (e.g., over-trust, shallow understanding, and reduced debugging practice), motivating a shift in assessment and instruction away from ``writing code from scratch'' as the primary evidence of learning \cite{prather2023we, becker2023programming}. This body of work has grown quickly, with recent SIGCSE syntheses cataloging common teaching practices (e.g., policy changes, redesigned assignments, and new tool-focused activities) as well as open research challenges \cite{prather2025beyond}.

A related line of work focuses on concrete pedagogical interventions for the ``AI era,'' including the design of new exercise formats that explicitly require students to reason about prompts, specifications, and verification rather than simply submit generated code \cite{denny2024promptproblems}, and large-scale course experiences that document how instructors message AI use, structure support, and adapt assessments \cite{liu2024cs50ai, fernandez2024cs1sideai}. Our emphasis on code comprehension, design discussion, and deliberate practice with large (and occasionally imperfect) codebases aligns with these efforts.

In parallel, curricula and degree programs are adapting to rapid automation by reconsidering which fundamentals should be explicitly taught and how the instructor's role changes when explanations and examples are instantly available. Kucera et al. describe this transition as requiring intentional curricular redesign and a reframing of the educator as a guide who supports judgment, ethics, and learning processes rather than acting solely as a source of technical answers \cite{Kucera2025Adapting}. Complementary work on AI-era assessment argues that evaluation must increasingly target process, explanation, and validity checks (not just final outputs), and outlines research agendas for doing so at scale \cite{weng2024assessment, marco2025redefining}.

Finally, our ``foundations first'' orientation draws on longer-standing traditions of hands-on and ``unplugged'' approaches that prioritize mental models over tooling. Bell et al.'s unplugged activities demonstrate that core computing ideas can be taught through physical and paper-based exercises, while Resnick argues for learning environments built around projects, peer interaction, and playful experimentation \cite{bell2009computer, resnick2017lifelong}. We combine these approaches with the realities of an adult, diverse student population in a bridge context \cite{Schmidt2025AnMS}, while also building on prior calls to treat code reading and discussion as first-class learning activities (e.g., pedagogical code reviews) \cite{hundhausen2013talking}.

\section{Course Design and Pedagogy}

\subsection{Schedule and Flipped Classroom Logistics}

To support a high-contact studio model, we consolidated the disparate schedules of CS1 and Discrete Math into a unified block. The primary studio session met every Tuesday from 10:45 AM to 3:00 PM. This schedule included a shared lunch break, which, while originally a logistical necessity, evolved into a valuable period for impromptu learning. Students often continued discussions over lunch, fostering a community of practice that is difficult to replicate in disjointed 50-minute lectures.

We also held weekly recitations on Thursdays (2 hours), which were used for guided exploration (e.g., Git and GitHub in Week 0) but remained less structured than the studio sessions.

Students prepared for the studio sessions by engaging with video content hosted on Canvas, linked from a custom course portal (\url{https://cs5001-5002.vankhoury.dev/}). We made a strategic decision not to film new lectures. Instead, we curated existing high-quality video content from standard CS 5001 and CS 5002 offerings. This allowed us to focus our development efforts entirely on the synchronous in-class experience and the design of new integrated projects. For the mathematical components, we are grateful to Dr. Richard Hoshino for sharing his problem sets, which we adapted from traditional homework assignments into collaborative in-class group exercises. To model the practices we taught, the authors collaborated on all course materials using a shared GitHub repository, following the same Git workflows required of students.

\subsection{The ``Studio'' Environment}

The 4-hour block allowed us to experiment with the flow between disciplines. Our initial ambition was to fluidly interweave CS1 and Discrete Math concepts throughout the session. In practice, the class settled into a natural rhythm: ``CS1 mode'' (programming concepts) typically dominated the morning session, while ``Discrete mode'' (mathematical theory) followed lunch.

However, the distinction between the two modes was intentionally blurred. ``CS1 mode'' often involved designing algorithms on whiteboards without computers, emphasizing logic over syntax. Conversely, ``Discrete mode'' frequently utilized Python frameworks to explore mathematical concepts computationally.

To support this integration, we meticulously merged the traditional syllabi of CS 5001 and CS 5002. Rather than teaching two distinct courses side-by-side, we reorganized topics to highlight their conceptual bridges. Approximately 80\% of the original content remained but was reordered to maximize synergy. Crucially, we observed that basic syntax-focused topics (such as `for` loops) required less instructional time than in traditional semesters. This efficiency allowed us to double down on our 'Foundations First' principle, dedicating more studio hours to the underlying logic and system design rather than syntax drilling. For instance, \textit{Recursion} (CS1 Week 5) was aligned with \textit{Mathematical Induction} (CS 5002 Module 7) to be taught in the same week, allowing students to see the recursive step as the computational equivalent of the inductive step. Similarly, Python's \textit{Lists and Sets} (CS1 Week 6) was moved to Week 3 to coincide with \textit{Set Theory} (CS 5002 Module 3), providing immediate practical application for abstract definitions. Some topics, like \textit{Object-Oriented Programming}, were moved significantly earlier (from Week 11 to Week 6) to provide a vocabulary for modeling complex mathematical structures like graphs and probability distributions later in the term. Table~\ref{tab:weekly_topics} illustrates the integrated schedule.

\begin{table*}[t]
  \caption{Weekly Topic Integration in CS 1.5}
  \label{tab:weekly_topics}
  \begin{tabular}{c p{4.5cm} p{4.5cm} p{5.5cm}}
    \toprule
    Week & CS1 Concept (Implementation) & Discrete Math Concept (Theory) & Integrated Theme / Bridge \\
    \midrule
    0 & Variables, Flowcharts & Number Bases (Binary, Hex) & Representation of Data \\
    1 & Conditionals, Booleans & Logic Gates, Truth Tables & Logic: Code $\leftrightarrow$ Math \\
    2 & Lists, Strings, Loops & Sets, Set Operations & Collections: Python \texttt{set} vs Math Set \\
    3 & Functions, Testing & Proofs, Cardinality & Specifications $\leftrightarrow$ Proofs; Testing $\leftrightarrow$ Verification \\
    4 & Recursion, While Loops & Mathematical Induction & Induction proves Recursion \\
    5 & Classes, Objects (OOP) & Combinatorics, Counting & Math objects (e.g. Fractions) as Classes \\
    6 & File I/O, Persistence & Probability, Random Variables & Monte Carlo Simulation \\
    7 & Stacks, Queues, Exceptions & (Review / Applied Probability) & Abstract Data Types \\
    8 & Graphs, BFS, DFS & Graph Theory, Traversal & Graph Algorithms \& Implementations \\
    9 & Linked Lists, Trees, BSTs & Trees, Tries & Hierarchical Structures \\
    10 & Searching, Sorting, Big O & Asymptotics, Complexity & Efficiency Analysis \\
    11 & Architecture (ECS), Crypto & Cryptography, Hashing & Applied Math: Security \& Systems \\
    \bottomrule
  \end{tabular}
\end{table*}

\subsection{Human Connection: The Sharing Circle}

In an era where technical answers are a prompt away, we prioritized human communication. We arranged the classroom in a circle to start each session. In these mandatory ``sharing circles,'' every student, TA, and instructor shared a meaningful story related to their learning that week. Participation was structured playfully: sometimes round-robin, sometimes ``popcorn'' style, and occasionally by shuffling nametags to encourage impersonation and humor.

Initially planned for 10 minutes, these sessions grew to occupy 20--30 minutes of our studio time. We allowed this expansion because it became a consistently engaged portion of the class; laptops were closed, phones were away, and students were fully present. Instructors—both active software developers—used this time as a pedagogical tool, sharing war stories from their current projects to align with student struggles. Students frequently requested to share their screens to demo side projects or ``fun finds.'' This culture of sharing extended beyond the classroom into a dedicated Slack workspace (chosen over academic tools for its ``industry vibe''), where follow-up discussions thrived. For instance, when a student shared a breakthrough about recursion during the circle, the conversation continued on Slack with peers sharing similar ``aha!'' moments and code snippets, reinforcing the community of practice.

The pedagogical impact was profound. For example, during a discussion on graph traversal, a student shared a story about navigating a corn maze with her son, effectively explaining Breadth-First Search (BFS) versus Depth-First Search (DFS) through lived experience. In Week 3, another student—a former Philosophy major—shared how encountering the keyword \texttt{self} in Python triggered an existential reflection on the nature of ``self,'' which led him to independently deduce the principles of Object-Oriented Programming. These moments of narrative connection anchored abstract technical concepts in reality.

\subsection{Embodied Learning}
We embraced ``silly'' physical exercises to build intuition before touching the keyboard. We used knit and duct tape to physically construct networks and simulate graph algorithms. We purchased wooden nesting dolls (Matryoshka dolls) and used them as props to teach recursion, physically demonstrating the ``base case.''

A standout activity was the ``dancing bubble sort,'' inspired by the AlgoRythmics group, who illustrate sorting algorithms through folk dance\footnote{AlgoRythmics: Bubble-sort with Hungarian (Csángó) folk dance. \url{https://www.youtube.com/watch?v=Iv3vgjM8Pv4}}. We played the video, played Hungarian folk music on Spotify, and distributed printed numbers to students who then performed the algorithm. Crucially, specific roles were assigned: one student counted swaps, while another counted comparisons. This physical enactment allowed the class to collectively deduce the $O(n^2)$ time complexity from their shared experience.

Students responded to these activities with zero resistance; in fact, the classroom often felt like a playful, primary-school-like atmosphere. We explicitly cultivated a culture of ``learning by play,'' drawing on the philosophy that humans learn best when the stakes are low and engagement is high~\cite{resnick2017lifelong}. This philosophy extended to our preference for analog tools: whenever possible, we prioritized paper tracing tables, whiteboard designs, and physical props over screens. This minimized digital distractions and maximized human-to-human collaboration, ensuring that when students finally opened their laptops, they had a robust mental model of the problem they needed to solve.

\section{Integrated Projects}

The merger of CS1 and Discrete Math allowed us to design projects that would be impossible in a siloed curriculum. Throughout the semester, we assigned five regular projects and one final capstone project. All projects were designed from scratch by the authors to ensure they spanned both courses.

A defining characteristic of these assignments was the \textit{volume of code}. In traditional CS1 courses, students often write small scripts from scratch (e.g., 50--100 lines). In contrast, we provided large, complex codebases averaging 2,000 to 3,000 lines. The students' task was to navigate the structure, identify the ``wiring,'' and implement specific classes or functions within an existing framework. This simulated real-world software engineering, where reading code is more common than writing it. It also introduced an authentic challenge: the provided code occasionally contained bugs. While fixing them was not mandatory, many students eagerly embraced the challenge, reporting their findings on Slack—a dynamic rarely seen in textbook assignments.

To support navigation through these substantial codebases, we mandated the use of AI tools from Week 0. All students were instructed to configure GitHub Copilot (free for students) and were given access to high-performance models (such as Claude Sonnet 4.5 and Opus 4.5) via university licenses. We explicitly trained them to use these tools for system comprehension—asking questions like ``explain how this class interacts with the UI''—rather than just code generation.

We present three examples of these integrated projects below.

\subsection{Set Theory and Modern Tooling}
\begin{figure}[h]
  \centering
  \includegraphics[width=\linewidth]{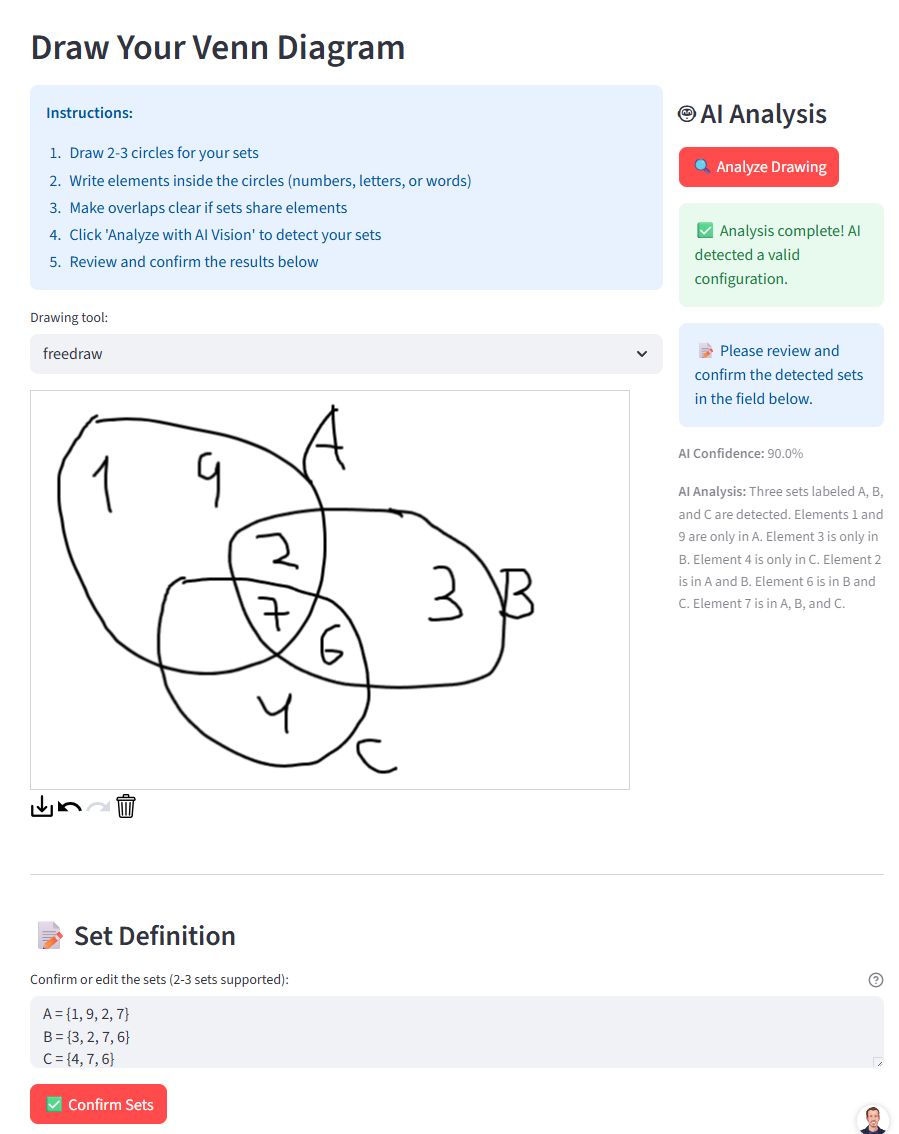}
  \caption{Set Theory Lab: Streamlit interface with Gemini-powered AI Assistant for diagram analysis.}
  \Description{Set Theory Lab application showing set operations and an AI chat assistant panel.}
  \label{fig:set_theory}
\end{figure}

To teach set operations, we assigned a project involving a codebase of roughly 1,800 lines that implemented a ``Set Theory Lab'' web application\footnote{Example implementation: \url{https://set-theory-lab.streamlit.app/}}. Students were tasked with implementing core set operations (such as symmetric difference and power sets) in the \texttt{src/} directory while navigating complex framework code (UI logic, configuration) that they did not write.

This project exposed students to a modern tech stack immediately. By design, each project introduced a new Python framework (\textit{Streamlit}\footnote{\url{https://streamlit.io/}}, \textit{Pygame}\footnote{\url{https://www.pygame.org/}}, or \textit{Flet}\footnote{\url{https://flet.dev/}}) to simulate the adaptability required in industry. For this assignment, we integrated \textit{OpenRouter}\footnote{\url{https://openrouter.ai/}}, providing students with sufficient funds to experiment with LLM-powered analysis features directly within their applications. A key feature involved students drawing set diagrams (Venn diagrams) on a web page canvas, and sending it to the multimodal Gemini 2.5 Flash model (via OpenRouter) to identify the sets and elements depicted. This feature seamlessly bridged physical intuition with AI-powered digital logic.

The integration of these tools served a distinct pedagogical goal: making the work ``portfolio-ready.'' We aimed to move beyond the disposable, terminal-based scripts typical of introductory courses, instead helping students build tangible, extendable applications they could proudly demonstrate to friends and family. To manage the complexity of the tech stack, students relied on the AI tools introduced in Week 0; prompts like ``explain this codebase'' allowed them to contribute to a system they did not fully understand—a common scenario in professional software engineering.

Pedagogically, the assignment enforced a feedback loop between math and code. For the CS 5002 (Math) component, students were required to choose test cases from the codebase's \texttt{doctests} and perform the actual calculations on paper, submitting scans of their work to verify that their code's logic matched mathematical reality.

\subsection{Recursion as a Game Mechanic}
\begin{figure}[h]
  \centering
  \includegraphics[width=\linewidth]{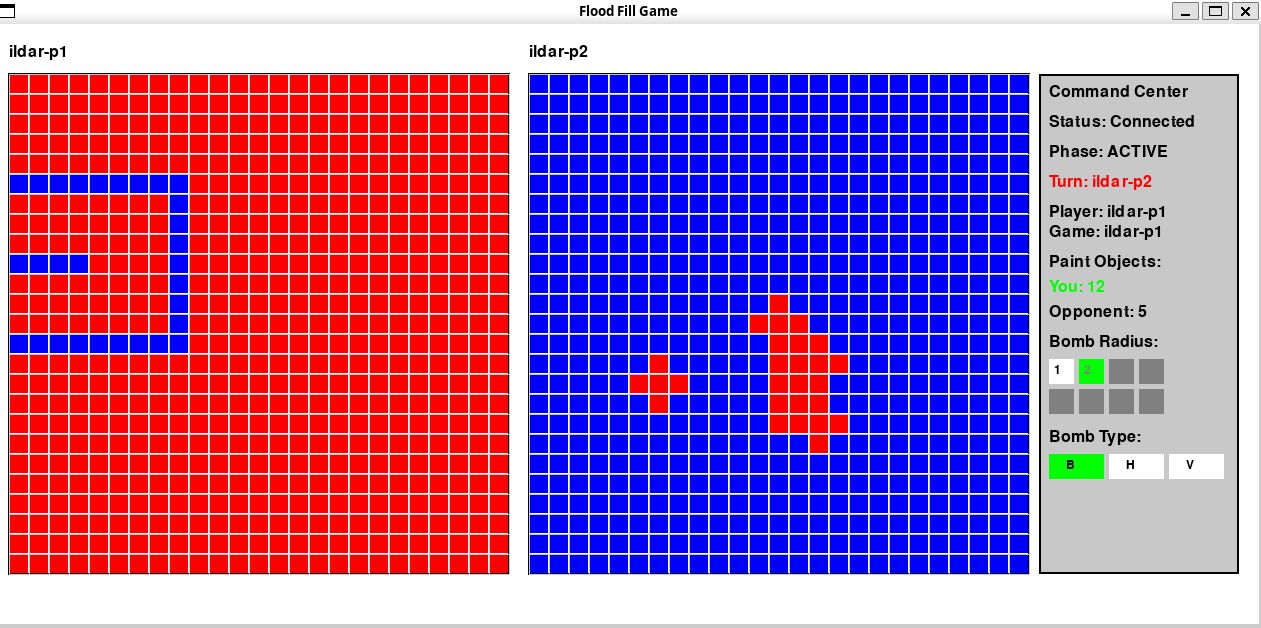}
  \caption{Recursive Flood Fill: A strategy game where move costs are derived via mathematical induction.}
  \Description{Game grid showing red and blue territories and bomb selection tools.}
  \label{fig:recursion_game}
\end{figure}

To bridge induction and recursion, we utilized a custom-built 2-player strategy game called "Recursive Flood Fill." Players competed to control territory on a 25x25 grid by placing "bombs" (Blob, Horizontal, or Vertical) that painted cells using recursive algorithms.

Crucially, the "cost" of using a bomb was not arbitrary; it was based on the mathematical formula for the number of cells covered by a given radius. In the \texttt{projected\_cost.py} file, students had to derive these formulas using mathematical induction. They were required to write out the Base Case, Inductive Hypothesis, and Inductive Step for each bomb type. Only after proving the formula could they implement the recursive flood fill logic in \texttt{paint.py}. To ensure rigor, a backend server (hosted on Render) verified the student's cost calculations against the ground truth, rejecting any moves where the implementation deviated from the proof. The project also included a "Go-like" capture mechanic requiring a Depth-First Search (DFS) implementation to detect surrounded regions, further reinforcing graph traversal concepts.

\subsection{Probability via Monte Carlo Simulation}
\begin{figure}[h]
  \centering
  \includegraphics[width=\linewidth]{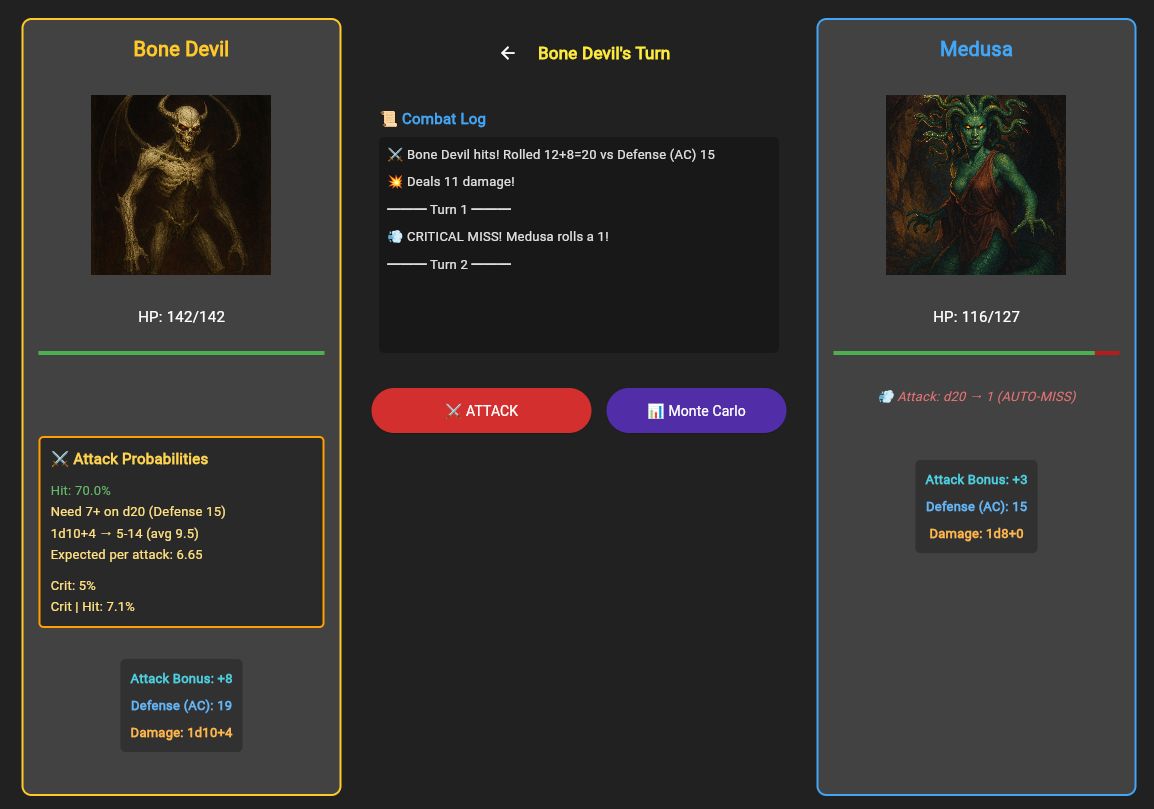}
  \caption{D\&D Combat Simulator: Flet app for empirically verifying probability calculations.}
  \Description{Simulator interface displaying monster configuration, controls, and convergence graphs.}
  \label{fig:dnd_sim}
\end{figure}

Probability is often abstract and disconnected from reality when taught via whiteboard formulas. To teach the Law of Large Numbers and Expected Value, we designed a Dungeons \& Dragons Combat Simulator\footnote{Example implementation: \url{https://cs1-discrete-project-dnd.onrender.com/}}. This project was structured in four phases: Randomness (dice rolling), Basic Probability, Conditional Probability, and Object-Oriented Testing. To enhance authenticity, the simulator fetched real monster statistics (Hit Points, Armor Class, etc.) from the D\&D 5th Edition REST API\footnote{\url{https://www.dnd5eapi.co/}}, requiring students to parse JSON responses and instantiate objects dynamically.

The assignment forced students to confront the relationship between theoretical probability and empirical results.
\begin{enumerate}
    \item \textbf{Manual Calculation:} Students selected five monster pairs (e.g., ``Glass Cannon vs. Tank'') and calculated win probabilities, expected damage, and conditional probabilities (P(crit|hit)) on paper.
    \item \textbf{Simulation:} Using the provided Monte Carlo framework, students ran simulations of 100, 1,000, and 5,000 battles.
    \item \textbf{Analysis:} Students submitted a report analyzing how the win rates ``stabilized'' as the sample size increased, effectively observing the Law of Large Numbers in action.
\end{enumerate}
If their simulation results for 5,000 battles did not match their paper math, they knew they had a bug—a lesson in verification that no textbook could provide. Interestingly, this rigorous verification process led students to discover a subtle bug in the instructor-provided codebase: a probability function was consistently yielding results slightly higher than the expected value derived on paper. This serendipitous discovery reinforced a critical lesson: no code is bug-free, and theoretical validation is an essential tool for debugging.

\section{Assessment: The Code Walk}

Recognizing that LLMs can generate correct solutions without student understanding, we fundamentally changed our assessment strategy. We instituted mandatory ``Code Walks'' for every project.
\begin{enumerate}
    \item \textbf{Peer Walk:} Students first explained their code line-by-line to a peer (not their project partner). They documented who they walked with and who walked them.
    \item \textbf{Instructor Walk:} Students then defended their implementation to a professor.
\end{enumerate}

Both peer and instructor walks were typically scheduled for 30 minutes. Unlike traditional academic defenses, we deliberately avoided rubrics to simulate industry code walkthroughs. Sessions typically began with an executive summary request: ``Explain your project as if we are your clients.'' After a live demonstration, the conversation shifted to the codebase.

We prioritized interactive, ``what-if'' scenarios over static explanations. Typical prompts included:
\begin{itemize}
    \item ``When you click this button, trace the execution path through the codebase.''
    \item ``How is this specific function tested? Show me the test case.''
    \item ``What happens if we change line $N$ to $X$? Let's make that change now and run it.''
\end{itemize}
This dynamic approach often led to live debugging sessions, turning assessment into a final teachable moment. We also seamlessly integrated the whiteboard; when tracing an execution path reached a mathematical concept (like a recursion cost or set operation), we would pause the code review and ask the student to derive the underlying formula on the board.

If a student could not explain a critical section of their code—a rare occurrence—they were asked to return the following week. We treated projects on a ``pass/fail'' basis to emphasize that in professional software engineering, a feature is either production-ready or it is not; there is no ``partial credit'' for broken code.

For the final project, every student participated in a one-hour code walk with both professors. This was resource-intensive but critical. It shifted the student's goal from ``getting it to work'' to ``understanding how it works.''

\section{Lessons Learned and Reflections}

\subsection{The Instructor as Mentor}
As outlined by Kucera et al., the era of automated software development demands that instructors evolve from content experts to mentors \cite{Kucera2025Adapting}. We found that our role shifted significantly from technical authorities to navigators of ambiguity. While AI could instantly generate a working binary search implementation, it could not help students decide \textit{whether} a binary search was the appropriate tool for a messy, real-world dataset. Our value lay in modeling how to decompose complex problems, how to verify AI-generated claims, and how to persist when a solution ``looked right'' but failed silently. The ``sharing circles'' and ``code walks'' became our primary teaching venues, allowing us to demonstrate professional judgment and ethical awareness—qualities that students absorb through human connection rather than prompt engineering. By relinquishing the need to be the source of all syntax knowledge, we were able to focus on the ``bigger picture'' skills that define a resilient computer scientist.

\subsection{Successes}
\begin{itemize}
    \item Rather than relying on traditional metrics like commit counts, we gauge success through longitudinal qualitative feedback. When this cohort transitioned to their subsequent semester, their new instructor reported an immediate difference: ``They are all friends, and they are fully engaged in class from day zero... I can do much more with this group.'' This suggests our emphasis on human connection fostered a lasting community of practice.
    \item The hypothesis that CS and Math are mirrors of each other was validated. Concepts like recursion/induction and objects/mathematical structures reinforced each other, leading to faster mastery of complex topics.
    \item By Week 12, students were not only comfortable navigating large, messy codebases but capable of architecting their own. For their final capstone, which had no starter code, students produced substantial applications. Surprisingly, despite our encouragement to use AI, nearly half the class chose to write their final projects entirely by hand. Both groups performed equally well in the rigorous final code walks, demonstrating that our approach supported diverse paths to proficiency: some through AI-accelerated development, others through traditional construction.
\end{itemize}

\subsection{Challenges and Future Work}
\begin{itemize}
    \item The co-teaching model and hour-long code walks were possible due to a relatively small class size. Scaling this to hundreds of students will require a robust training program for TAs to conduct high-quality code walks and facilitate sharing circles. Finally, the small cohort size ($n=11$) limits the generalizability of our findings to larger populations without further study.
    \item The "intensive" nature of the combined course was truly intense. While the back-to-back scheduling provided deep immersion, it also tested the stamina of both students and instructors.
    \item Replicating this requires instructors who are willing to relinquish the "sage on the stage" persona and embrace a chaotic, experimental, and vulnerable teaching style.
    \item In our commitment to an experimental, adaptable style, we sometimes sacrificed administrative rigor. Ad hoc processes—particularly for scheduling code walks—were a point of friction noted in student evaluations. Future iterations will require a more structured administrative backbone to support the pedagogical flexibility.
\end{itemize}

\section{Conclusion}
``CS 1.5'' was a risky experiment, but a necessary one. We successfully demonstrated that by embracing AI, prioritizing human connection, and explicitly bridging mathematical theory with computational practice (e.g., linking induction to recursion), we can create a curriculum that is better adapted to the realities of the AI era. We invite other institutions to join us in reimagining the foundations of computer science education—not as a transfer of syntax, but as a cultivation of intuition, logic, and professional judgment. To facilitate this transition, all project codebases and assignment materials are available as open-source resources\footnote{\url{https://github.com/cs1-5}}, and we welcome contributions, adaptations, and forks from educators who wish to build on or remix this curriculum for their own contexts.

\section{Acknowledgments}
We thank the Khoury College of Computer Sciences at Northeastern University for supporting this experiment. We are deeply indebted to the 11 students of the Fall 2025 cohort, whose enthusiasm, trust, and willingness to embrace the unknown made this experiment possible.

\bibliographystyle{ACM-Reference-Format}
\balance
\bibliography{references} 

@inproceedings{Kucera2025Adapting,
author = {Kucera, Ludek and Akhmetov, Ildar and Izu, Cruz and Omojokun, Olufisayo},
title = {Adapting Computing Curricula in the Era of Automated Software Development},
year = {2025},
isbn = {9798400719424},
publisher = {Association for Computing Machinery},
address = {New York, NY, USA},
url = {https://doi.org/10.1145/3736251.3754332},
doi = {10.1145/3736251.3754332},
booktitle = {Proceedings of the ACM Global on Computing Education Conference 2025 Vol 2},
pages = {353–355},
numpages = {3},
location = {Gaborone, Botswana},
series = {CompEd 2025}
}

@book{resnick2017lifelong,
  title={Lifelong Kindergarten: Cultivating Creativity through Projects, Passion, Peers, and Play},
  author={Resnick, Mitchel},
  year={2017},
  publisher={MIT Press}
}

@book{bell2009computer,
  title={Computer Science Unplugged: School Students Doing Real Computing Without Computers},
  author={Bell, Tim and Alexander, Jason and Freeman, Isaac and Grimley, Mick},
  journal={New Zealand Journal of applied computing and information technology},
  volume={13},
  number={1},
  pages={20--29},
  year={2009}
}

@inproceedings{Schmidt2025AnMS,
author = {Schmidt, Logan W. and Kidder, Caitlin J. and Akhmetov, Ildar and Bebis, Megan and Jamieson, Alan C. and Lionelle, Albert and Maravetz, Sarah and Rollins, Sami and Selinger, Ethan},
title = {An MS in CS for non-CS Majors: A Ten Year Retrospective},
year = {2025},
isbn = {9798400705311},
publisher = {Association for Computing Machinery},
address = {New York, NY, USA},
url = {https://doi.org/10.1145/3641554.3701928},
doi = {10.1145/3641554.3701928},
booktitle = {Proceedings of the 56th ACM Technical Symposium on Computer Science Education V. 1},
pages = {1036–1042},
numpages = {7},
location = {Pittsburgh, PA, USA},
series = {SIGCSETS 2025}
}

@article{prather2023we,
  title={"It's Weird That it Knows What I Want": Usability and Interactions with Copilot for Novice Programmers},
  author={Prather, James and Reeves, Brent N. and Denny, Paul and Becker, Brett A. and Leinonen, Juho and Luxton-Reilly, Andrew and Powell, Garrett B. and Finnie-Ansley, James and Santos, Eddie Antonio},
  journal={ACM Transactions on Computer-Human Interaction},
  volume={31},
  pages={1--31},
  year={2023},
  doi={10.1145/3617367}
}

@inproceedings{becker2023programming,
  title={Programming Is Hard-Or at Least It Used to Be: Educational Opportunities and Challenges of AI Code Generation},
  author={Becker, Brett A and Denny, Paul and Finnie-Ansley, James and Luxton-Reilly, Andrew and Prather, James and Santos, Eddie Antonio},
  booktitle={Proceedings of the 54th ACM Technical Symposium on Computer Science Education V. 1},
  pages={500--506},
  year={2023}
}

@article{prather2025beyond,
  title={Beyond the hype: A comprehensive review of current trends in generative AI research, teaching practices, and tools},
  author={Prather, James and Leinonen, Juho and Kiesler, Natalie and Gorson Benario, Jamie and Lau, Sam and MacNeil, Stephen and Norouzi, Narges and Opel, Simone and Pettit, Vee and Porter, Leo and others},
  journal={2024 Working Group Reports on Innovation and Technology in Computer Science Education},
  pages={300--338},
  year={2025}
}

@inproceedings{denny2024promptproblems,
  title={Prompt Problems: A new programming exercise for the generative AI era},
  author={Denny, Paul and Leinonen, Juho and Prather, James and Luxton-Reilly, Andrew and Amarouche, Thezyrie and Becker, Brett A and Reeves, Brent N},
  booktitle={Proceedings of the 55th ACM Technical Symposium on Computer Science Education V. 1},
  pages={296--302},
  year={2024}
}

@inproceedings{liu2024cs50ai,
  title={Teaching CS50 with AI: leveraging generative artificial intelligence in computer science education},
  author={Liu, Rongxin and Zenke, Carter and Liu, Charlie and Holmes, Andrew and Thornton, Patrick and Malan, David J},
  booktitle={Proceedings of the 55th ACM technical symposium on computer science education V. 1},
  pages={750--756},
  year={2024}
}

@inproceedings{fernandez2024cs1sideai,
author = {Fernandez, Amanda S. and Cornell, Kimberly A.},
title = {CS1 with a Side of AI: Teaching Software Verification for Secure Code in the Era of Generative AI},
year = {2024},
isbn = {9798400704239},
publisher = {Association for Computing Machinery},
address = {New York, NY, USA},
url = {https://doi.org/10.1145/3626252.3630817},
doi = {10.1145/3626252.3630817},
booktitle = {Proceedings of the 55th ACM Technical Symposium on Computer Science Education V. 1},
pages = {345–351},
numpages = {7},
location = {Portland, OR, USA},
series = {SIGCSE 2024}
}

@article{weng2024assessment,
  title={Assessment and learning outcomes for generative AI in higher education: A scoping review on current research status and trends},
  author={Weng, Xiaojing and Qi, XIA and Gu, Mingyue and Rajaram, Kumaran and Chiu, Thomas KF},
  journal={Australasian Journal of Educational Technology},
  year={2024}
}

@inproceedings{marco2025redefining,
  title={Redefining programming assessments in the era of artificial intelligence and large language models},
  author={Marco-Detchart, C and Lopez-Molina, C},
  booktitle={INTED2025 Proceedings},
  pages={5610--5615},
  year={2025},
  organization={IATED}
}

@article{hundhausen2013talking,
  title={Talking about code: Integrating pedagogical code reviews into early computing courses},
  author={Hundhausen, Christopher D and Agrawal, Anukrati and Agarwal, Pawan},
  journal={ACM Transactions on Computing Education (TOCE)},
  volume={13},
  number={3},
  pages={1--28},
  year={2013},
  publisher={ACM New York, NY, USA}
}

\end{document}